\documentclass[fleqn,usenatbib]{mnras}

\usepackage[T1]{fontenc}

\DeclareRobustCommand{\VAN}[3]{#2}
\let\VANthebibliography\thebibliography
\def\thebibliography{\DeclareRobustCommand{\VAN}[3]{##3}\VANthebibliography}

\usepackage{graphicx}	% Including figure files
\usepackage{amsmath}	% Advanced maths commands
\usepackage{amssymb}	% Extra maths symbols
\usepackage{multirow}
\usepackage{booktabs}
\usepackage{mhchem}
\usepackage{tikz,xcolor,hyperref}
\usepackage{orcidlink}
\makeatletter
\newcommand*{\linktocite}[2]{%
  \hyper@natlinkstart{#1}#2\hyper@natlinkend}
\makeatother

\title[\textit{JWST} Predictions for LP~890-9~c]{A Venus in the Making? Predictions for \textit{JWST} Observations of the Ultracool M-Dwarf Planet LP~890-9~c}

\author[Gomez Barrientos et al.]{
Jonathan Gomez Barrientos \orcidlink{0000-0002-0672-9658}$^{1,2}$\thanks{E-mail: jgomezba@caltech.edu},
Lisa Kaltenegger \orcidlink{0000-0002-0436-1802}$^{1}$,
and Ryan J. MacDonald \orcidlink{0000-0003-4816-3469}$^{1,3,4}$
\\
$^{1}$Carl Sagan Institute and Department of Astronomy, Cornell University, Ithaca, NY 14853, USA \\
$^{2}$Division of Geological and Planetary Sciences, California Institute of Technology, Pasadena, CA 91125, USA \\
$^{3}$Department of Astronomy, University of Michigan, Ann Arbor, MI 48109, USA \\
$^{4}$NHFP Sagan Fellow \\
}

\date{Accepted 2023 May 10. Received 2023 April 25; in original form 2022 December 23}

\pubyear{2023}

\begin{document}
\label{firstpage}
\pagerange{\pageref{firstpage}--\pageref{lastpage}}
\maketitle

% Abstract of the paper
\begin{abstract} 

\noindent The recently discovered transiting super-Earth LP~890-9~c is potentially one of the best rocky exoplanets for atmospheric characterization. Orbiting an ultracool M-dwarf at the inner edge of the habitable zone, LP~890-9~c offers a new opportunity to study the climate of rocky planets at the inner edge of the habitable zone. We investigate the molecular detectability with simulated \textit{JWST} transmission spectra for five potential atmospheres of LP~890-9~c. We find that a small three-transit \textit{JWST} program can infer evidence of $\ce{H2O}$ (at 3$\sigma$ confidence) for a full runaway greenhouse scenario. Alternatively, $\ce{CO2}$-dominated atmospheres resembling Venus without high-altitude terminator clouds can be identified with eight transits. However, these predictions could be complicated by the impact of clouds and/or unocculted starspots. Nevertheless, \textit{JWST} observations of LP~890-9~c could provide critical insights and potentially distinguish between models of rocky planets at the inner edge of the habitable zone.

\end{abstract}

% Select between one and six entries from the list of approved keywords.
\begin{keywords}
planets and satellites: terrestrial planets -- planets and satellites: individual (LP~890-9~c) -- planets and satellites: atmospheres -- techniques: spectroscopic
\end{keywords}

%%%%%%%%%%%%%%%%%%%%%%%%%%%%%%%%%%%%%%%%%%%%%%%%%%

%%%%%%%%%%%%%%%%% BODY OF PAPER %%%%%%%%%%%%%%%%%%

\section{Introduction} \label{sec:intro}

Planets orbiting ultra-cool M dwarf stars present one of the best opportunities to characterize terrestrial exoplanet atmospheres. The best known example, the TRAPPIST-1 system \citep{gillon2016, gillon2017}, has ignited intense theoretical and observational study since its discovery in 2016. With \textit{JWST} fully operational, the atmospheres of these rare ultra-cool M dwarf planets can now be explored.

Recently, \citet{Delrez2022} reported the discovery of two transiting super-Earths around the low-activity M6 dwarf LP~890-9 (also known as TOI-4306 or SPECULOOS-2). The second planet in the system, LP~890-9~c --- discovered by the ground-based SPECULOOS survey \citep{gillon2018,sebastian2021} --- has a radius of $1.367^{+0.055}_{-0.039}$~$\textup{R}_{\oplus}$, an orbital period of 8.46 days, and receives slightly less irradiation than Earth (0.906~$\pm$ 0.026~$\textup{S}_{\oplus}$). Since M dwarfs warm Earth-like planets more effectively than Sun-like stars (see, e.g., \citealt{kasting1993}), LP~890-9~c lies very close to the inner edge of the conservative habitable zone (HZ) \citep{kopparapu2013, kopparapu2017}. After the TRAPPIST-1 planets, LP~890-9~c is the next best HZ terrestrial planet for atmospheric characterization with \textit{JWST} \citep{Delrez2022}, while also providing initial insights into the conditions of a potentially habitable world at the inner edge of the conservative HZ.

Recently, \citet{Kaltenegger2022} presented initial atmospheric models for five possible atmospheres of LP~890-9~c: (1) a Hot Earth analogue; (2) a moist greenhouse with the same $\ce{CO2}$ abundance as the modern Earth; (3) a moist Runaway Greenhouse with 20$\%$ of the $\ce{CO2}$ abundance of the dry atmosphere; (4) a Full Runaway Greenhouse with a steam atmosphere; and (5) a modern Venus analogue dominated by $\ce{CO2}$. In what follows, we refer to these five models as `Hot Earth', `Runaway 1,2,3', and `Exo-Venus, cloud-free'. The Hot Earth scenarios uses modern Earth models at the orbital distance of LP890-9c. Runaway 1 and 2 show increasing surface temperature ($T_{\rm{surf}}$) and increasing water vapor in the atmosphere, which in turn increases the total surface pressure ($P_{\rm{surf}}$) as expected for terrestrial planet atmospheres near the inner edge of the HZ, leading to a $P_{\rm{surf}}$ of 1.27 bar and 3.86 bar for these scenarios. Runaway 3 shows a full runaway greenhouse steam atmosphere, with a $P_{\rm{surf}}$ of 271 bar, and $T_{\rm{surf}}$ of 1600 K. The Exo-Venus scenario uses the $\ce{CO2}$ dominated atmospheric model with $P_{\rm{surf}}$ of 153 bar and $T_{\rm{surf}}$ of 599~K. All five models assume no clouds at the terminator region. To explore the effect of Venus-like clouds, we also consider an opaque cloud layer at 70km for the Exo-Venus scenario.

In this Letter, we evaluate the ability of \textit{JWST} to characterize the atmosphere of LP~890-9~c. Since planets at the inner edge of the conservative HZ like LP~890-9~c can sustain several very different possible atmospheres (see e.g., \citealt{fauchez2022,turbet2020,Kaltenegger2022}), depending on the initial water inventory and the water loss timescales, we consider the five scenarios described above. We perform an atmospheric retrieval analysis for these possible atmospheres of LP~890-9~c, which could represent steps in the evolution of a terrestrial planet from an Earth-like planet to one resembling Venus \citep{Kaltenegger2022}. Our retrieval analysis quantifies the molecular detectability and anticipated abundance constraints for each LP~890-9~c atmospheric scenario, demonstrating the priority of observing this tantalizing planet with \textit{JWST}.

\section{Methods}
\label{sec:methods}

\subsection{Model Spectra of LP~890-9~c}
\label{sec:models}

We model and retrieve transmission spectra for five atmospheres for LP~890-9~c \citep{Kaltenegger2022} with the radiative transfer and retrieval code POSEIDON \citep{MacDonald2017,MacDonald2023}. First, we use POSEIDON's forward model, TRIDENT \citep{MacDonald2022}, to compute transmission spectra from the 1D atmospheric pressure-temperature (P-T) and chemical abundance profiles from \citet{Kaltenegger2022}. We interpolated these profiles onto a 100-layer vertical grid evenly spaced in log-pressure from $10^{-7}$--$10$\,bar.  We assign the planet's white light radius (1.37\,$\textup{R}_{\oplus}$) to the surface and compute a new reference radius at a pressure of 1~mbar and, in lieu of a mass measurement, adopt the predicted median mass of 2.5,$M_{\oplus}$\citep{Delrez2022} from a mass-radius relationship \citep{Chen2017}. For the radiative transfer calculation, our models consider $\ce{N2, H2O}$ and $\ce{CO2}$ (following \citealt{kopparapu2013}), with the exception of the Hot Earth scenario that additionally includes $\ce{O2, O3, CH4}$ and $\ce{N2O}$. The opacity data used by TRIDENT are described in \citet{MacDonald2022} (their Appendix C). 

Since the anticipated cloud pressure level for LP~890-9~c is uncertain, in this initial study we generally solve for the radiative transfer assuming clear atmospheres. However, we briefly explore the effect of clouds by adding an opaque surface, concealing layers deeper than 70\,km altitude, to the Exo-Venus model. Our model transmission spectra are shown in Figure~\ref{fig:models_and_data}.

\subsection{Simulated \textit{JWST} Observations}
\label{sec:observations}

To assess the number of transits of LP~890-9~c required to detect key molecular species, we next generated simulated \textit{JWST} data using PandExo \citep{Batalha2017}. We considered 1 to 200 transits to generate synthetic \textit{JWST} observations, depending on the atmospheric scenario, to investigate how many transits are required to achieve at least a 3 sigma detection for a given molecule. We initially considered both NIRSpec PRISM (0.6--5.3\,$\mu$m) and MIRI LRS (5--12\,$\mu$m) observations to explore the full wavelength coverage of our model transmission spectra. For NIRSpec PRISM, we use the 512 subarray. For both instruments, we limit the saturation to 80$\%$ full well. We keep all simulated observations at their native resolution for retrieval purposes to avoid losing information by binning to a specific spectral resolution. We use the stellar spectrum from \citet{Delrez2022}, normalized to a magnitude of J = 12.258. Finally, we assume an observation baseline for each transit of three times the transit duration (3 $\times$ 57.56\,min; \citealt{Delrez2022}).

\begin{figure}
    \centering
    \includegraphics[width=\columnwidth]{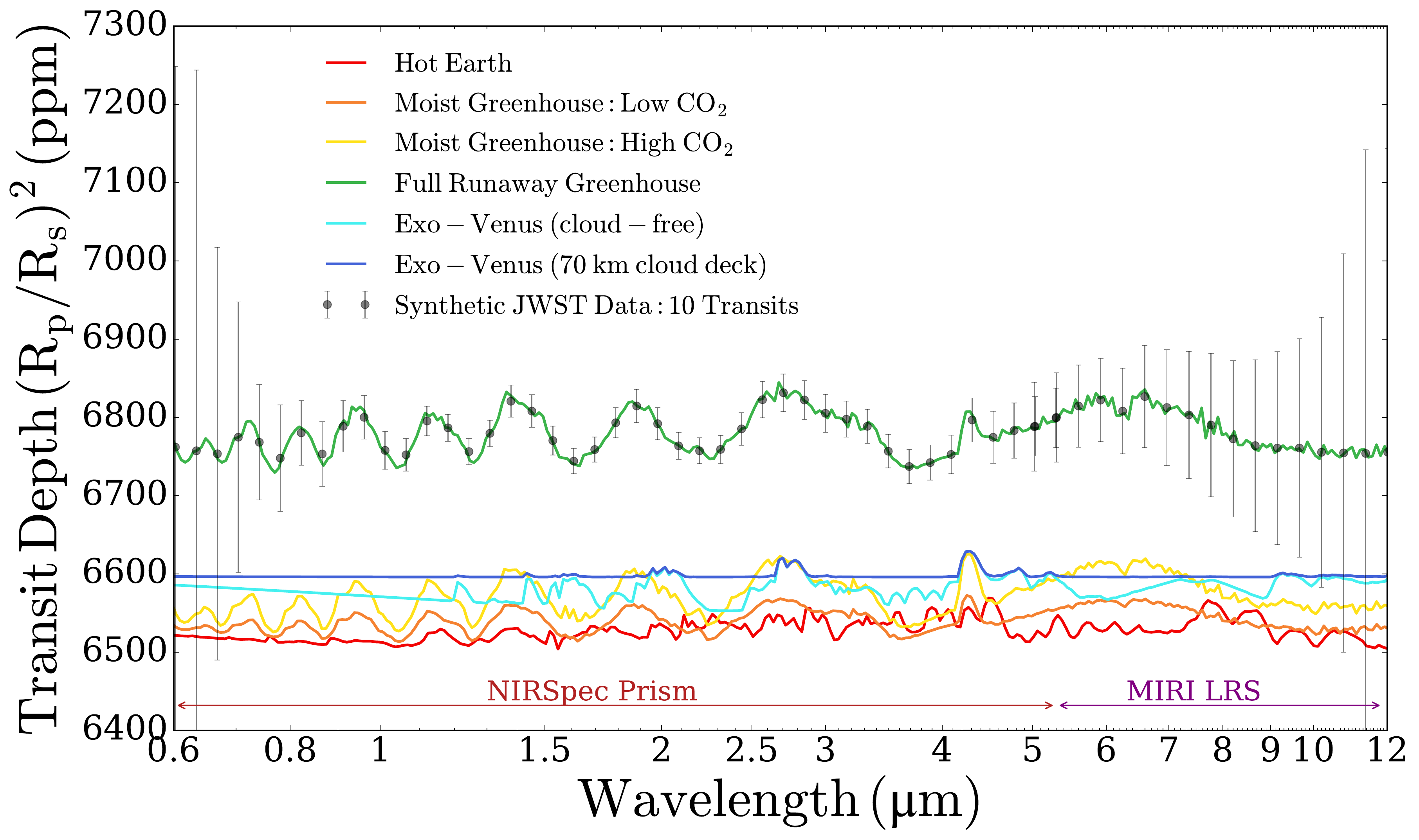}
    \vspace{-0.2cm}
    \caption{Model transmission spectra and simulated \textit{JWST} observations of LP~890-9~c. The atmospheric scenarios from \citet{Kaltenegger2022} (colored lines) are shown binned to a spectral resolution of $R = 100$. We overplot \textit{JWST} NIRSpec PRISM and MIRI LRS simulations, assuming 10 transits with each mode, without Gaussian scatter, and binned to $R = 20$, for the Full Runaway Greenhouse model (Runaway 3) (the error bars have the same magnitude for the other scenarios). The transit depth offset between the Runaway 3 model and the other scenarios arises from the extended $\ce{H2O}$-rich atmosphere (substantially increasing the 1~mbar planet radius).}
    \label{fig:models_and_data}
\end{figure}

Figure~\ref{fig:models_and_data} shows our simulated \textit{JWST} observations for 10 transits with NIRSpec PRISM and 10 transits with MIRI LRS. Given the large error bars of the MIRI observations, and noting that previous retrieval studies have found that NIRSpec PRISM has the greatest information content \citep[e.g.][]{Batalha2019,Lin2021}, we only include the NIRSpec PRISM observations in our subsequent retrievals. For our retrieval analysis, we omit Gaussian scatter from the synthetic data and instead center the data on its (true) binned model location --- this essentially produces a mean retrieval result unaffected by a specific random noise draw \citep[see][]{Feng2018}.

\subsection{Atmospheric Retrieval Analysis}
\label{sec:retrievals}

We use the atmospheric retrieval code POSEIDON \citep{MacDonald2017}, which couples a parametric form of the transmission spectrum model described above with the nested sampling algorithm PyMultiNest \citep{Feroz2009, Buchner2014}, to establish predicted constraints on LP~890-9~c's atmosphere. We divide the atmosphere of LP~890-9~c into 100 layers equally spaced in log-pressure from $10^{-7}$--$10$\,bar. We parameterize the P-T profile as a free gradient (linear in log-pressure) between $10^{-5}$ and $1$\,bar \citep{MacDonald2022}, described by the temperatures at these boundary pressures (each with a uniform prior from 100--800\,K). Two parameters are assigned for the 1~mbar reference radius of the planet (uniform prior from 0.8--1.2\,$\mathrm{R_p}$) and for the surface/cloud pressure (log-uniform prior from $10^{-7}$--$10$\,bar). Since the molecule dominating LP~890-9~c's atmosphere is \emph{a priori} unknown, and differs across our five atmosphere scenarios, we adopt the centered log-ratio (CLR) of the gases as free parameters \citep[e.g.][]{Aitchison1986,Benneke2012}. The CLR parameterization treats all gases with equal priors, while automatically accounting for the summation to unity constraint of mixing ratios \citep{Benneke2012}. For the Runaway Greenhouse and Exo-Venus models, we fit for the CLR-transformed mixing ratios of $\ce{H2, H2O, and CO2}$. For the Hot Earth scenario, we fit six parameters for $\ce{O2, O3, H2O, CO2, CH4, and N2O}$. The CLR priors span the parameter space from a minimum mixing ratio of $10^{-12}$ to any of the gases dominating the atmosphere (see \linktocite{Lustig-YaegerFu2023}{Lustig-Yaeger \& Fu et al.} \citeyear{Lustig-YaegerFu2023}). We fill the remaining atmosphere with $\ce{N2}$ (specified by the summation to unity condition without requiring a free parameter). Thus, depending on the atmospheric scenario, our retrievals contain a total of 7--10 free parameters. All our retrievals sample the parameter space with 2000 MultiNest live points.

We conducted a total of 54 retrievals. For each simulated dataset, we perform a retrieval with the full reference set of parameters described above. We then computed additional nested retrievals with a single gas removed. By comparing the Bayesian evidences of these models, we compute predicted detection significances of $\ce{H2O, CO2, CH4, and N2O}$ as a function of the number of transits through Bayesian model comparisons \citep[see e.g.,][]{Benneke2013, Trotta2017}.

\section{Results: \textit{JWST} Predictions for LP~890-9~c}
\label{sec:results}

\begin{figure}
    \centering
    \includegraphics[width=\columnwidth]{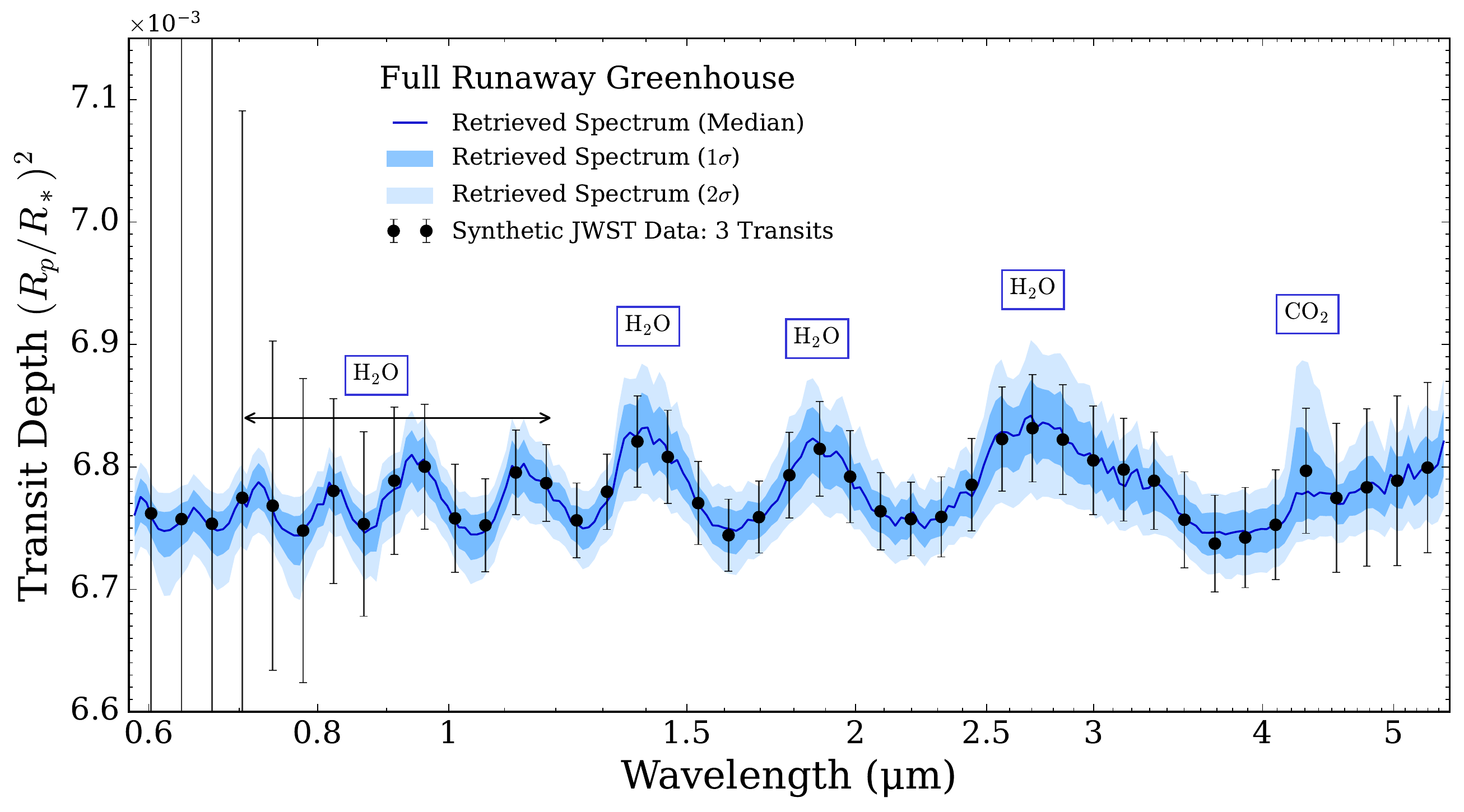}
    \includegraphics[width=\columnwidth]{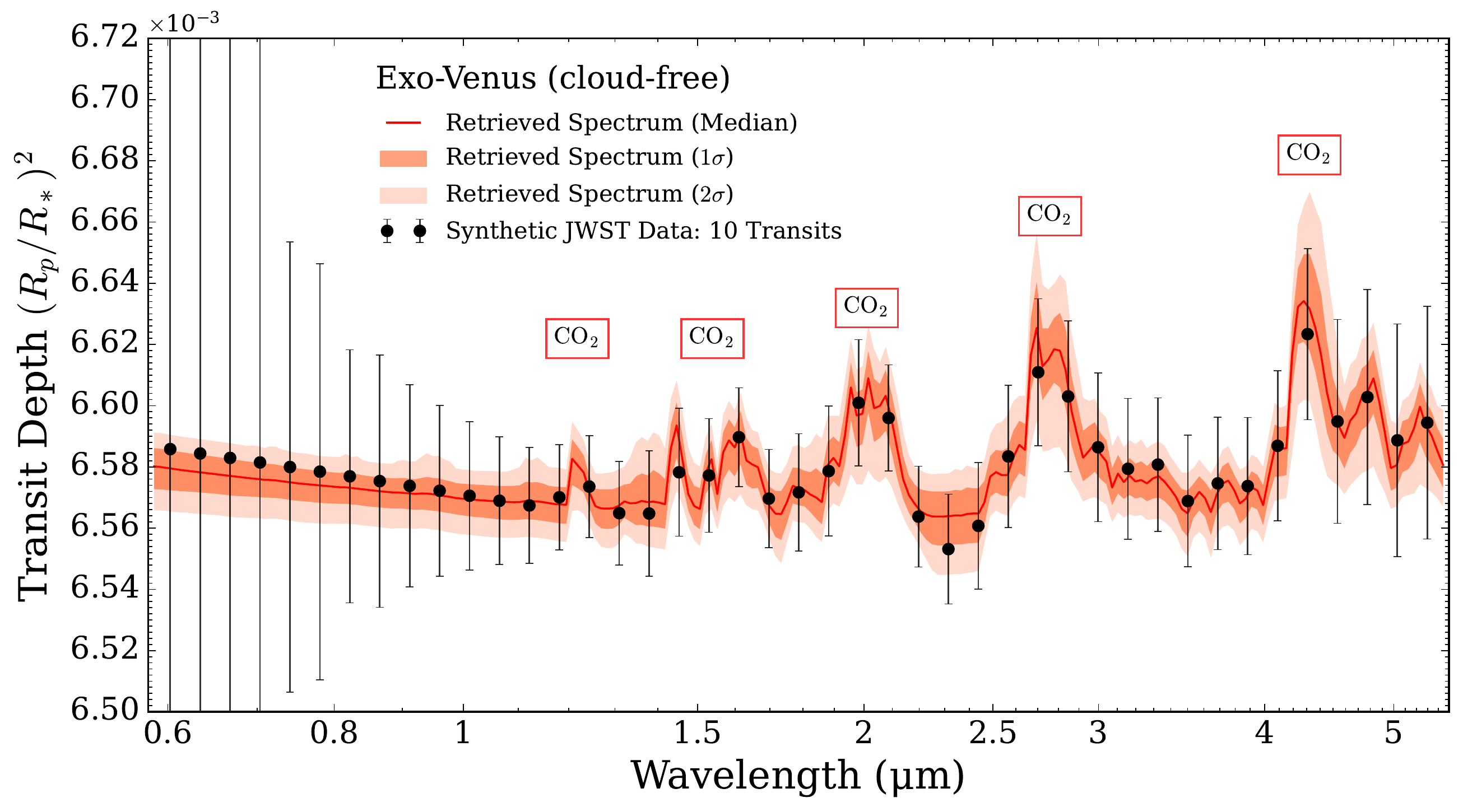}
    \includegraphics[width=\columnwidth]{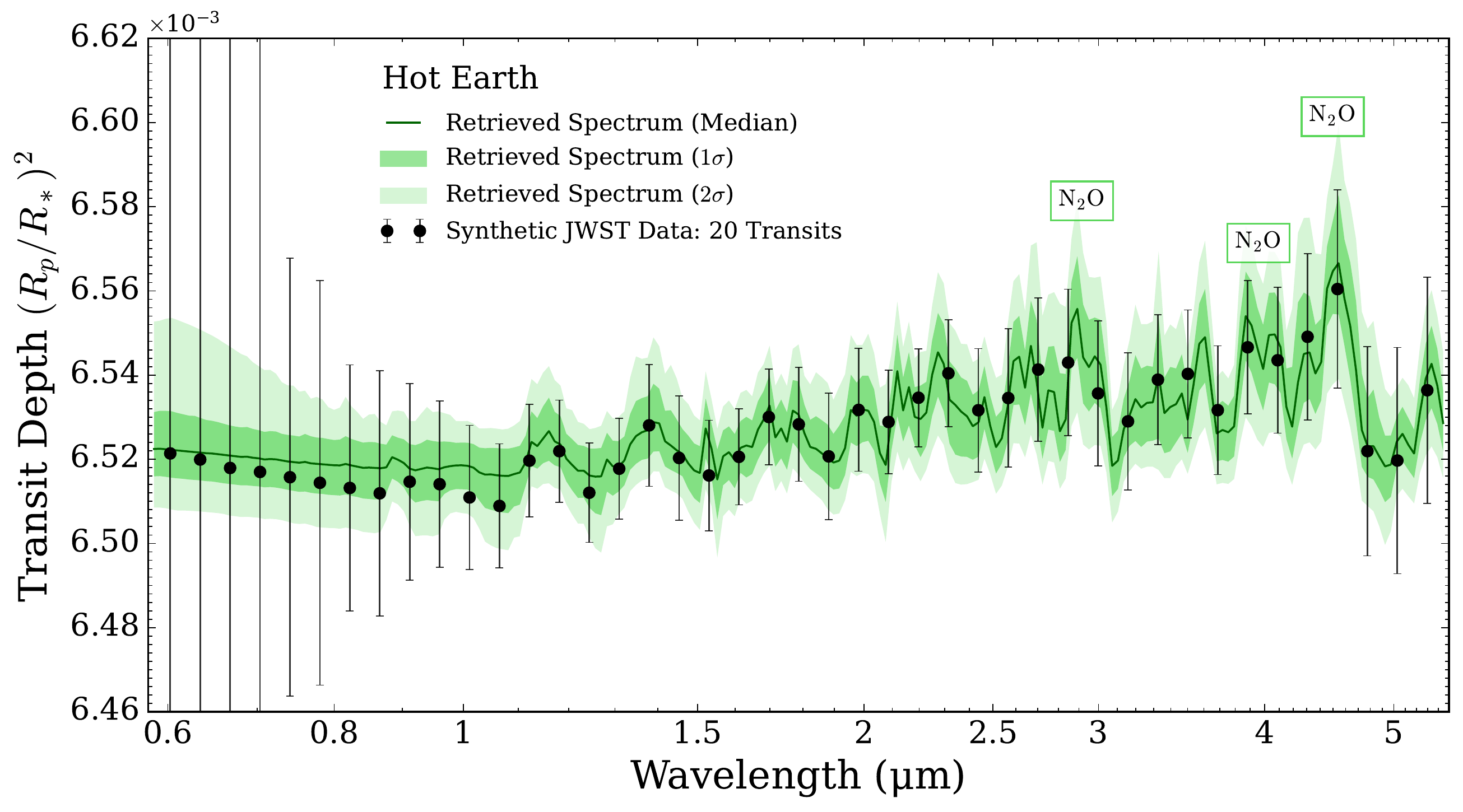}
    \caption{Retrieved transmission spectra from simulated \textit{JWST} observations of LP~890-9~c. \textit{Top:} Three synthetic NIRSpec PRISM observations of the 5 bar Full Runaway Greenhouse scenario (Runaway 3). \textit{Middle:} Ten synthetic NIRSpec PRISM observations of the 5 bar cloud-free Exo-Venus scenario. \textit{Bottom:} Twenty synthetic NIRSPEC PRISM observations of the 1 bar Hot Earth scenario. The median retrieved transmission spectra (solid line) and the 1$\sigma$/2$\sigma$ (shaded regions) are shown. The most prominent spectral features are labeled. All datasets are shown binned to a spectral resolution of R=20 for clarity and without Gaussian scatter.}
    \label{fig:retrieved_spectra}
\end{figure}

Our transmission spectra show marked differences between the model atmospheres (Figure \ref{fig:models_and_data}, see also \citealt{Kaltenegger2022}). Hence, we find that even moderately sized \textit{JWST} programs can distinguish between different atmospheric scenarios that are crucial for understanding the nature of planets near the inner edge of the HZ. Figure~\ref{fig:retrieved_spectra} demonstrates this for two such scenarios: a $\ce{H2O}$-dominated Full Runaway Greenhouse (Runaway 3), a $\ce{CO2}$-dominated cloud-free Exo-Venus. In each case, our retrievals demonstrate that $<$ 10 transits with \textit{JWST} could establish whether LP~890-9~c's atmosphere is dominated by $\ce{H2O}$ or $\ce{CO2}$. The third case shown in 
Figure~\ref{fig:retrieved_spectra} shows that 20 \textit{JWST} transits could identify $\ce{N2O}$ in the atmosphere of a Hot Earth scenario, which could provide a stable habitable environment for LP~890-9~c \citep{Kaltenegger2022}.

\begin{figure}
    \centering
    \includegraphics[width=\columnwidth, trim = 0.6cm 0.2cm 0.6cm 0.2cm]{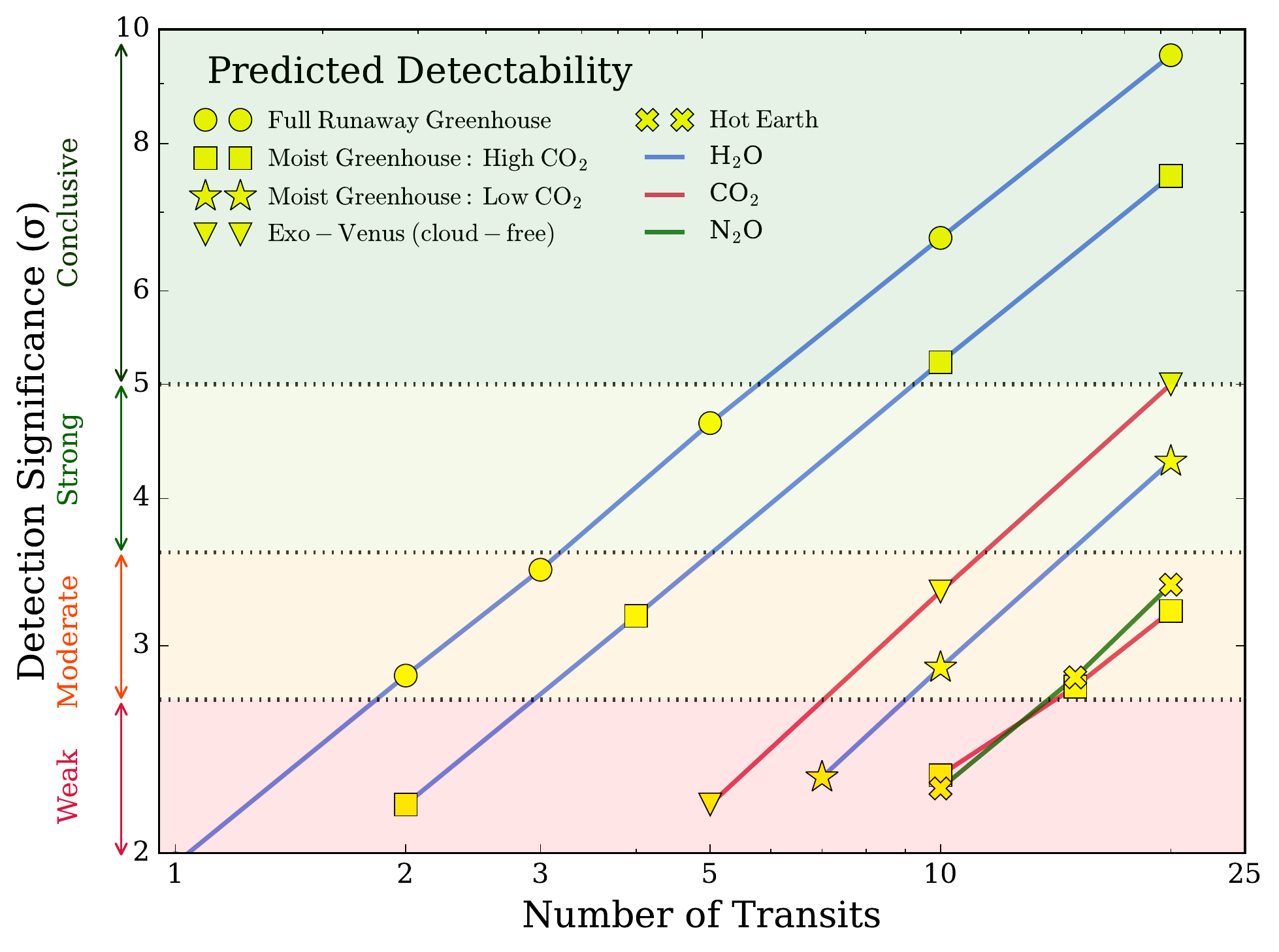}
    \vspace{-0.2cm}
    \caption{Predicted molecular detection significances for five possible LP~890-9~c atmosphere models as a function of \textit{JWST} NIRSpec PRISM transits. The detection significances are categorized as ‘weak’, ‘moderate’, ‘strong’, and ‘conclusive’ detections (colored regions), according to an adaptation of the Jeffreys’ scale for Bayesian model comparisons \citep{Trotta2017}. The boundaries (dotted lines) between the different categories occur at $2.7\sigma$, $3.6\sigma$, and $5.0\sigma$. The marker symbols correspond to each atmospheric scenario, while the line colors corresponds to each molecule.} 
    \label{fig:detections}
\end{figure}

\begin{figure*}
    \centering
    \includegraphics[width=\textwidth]{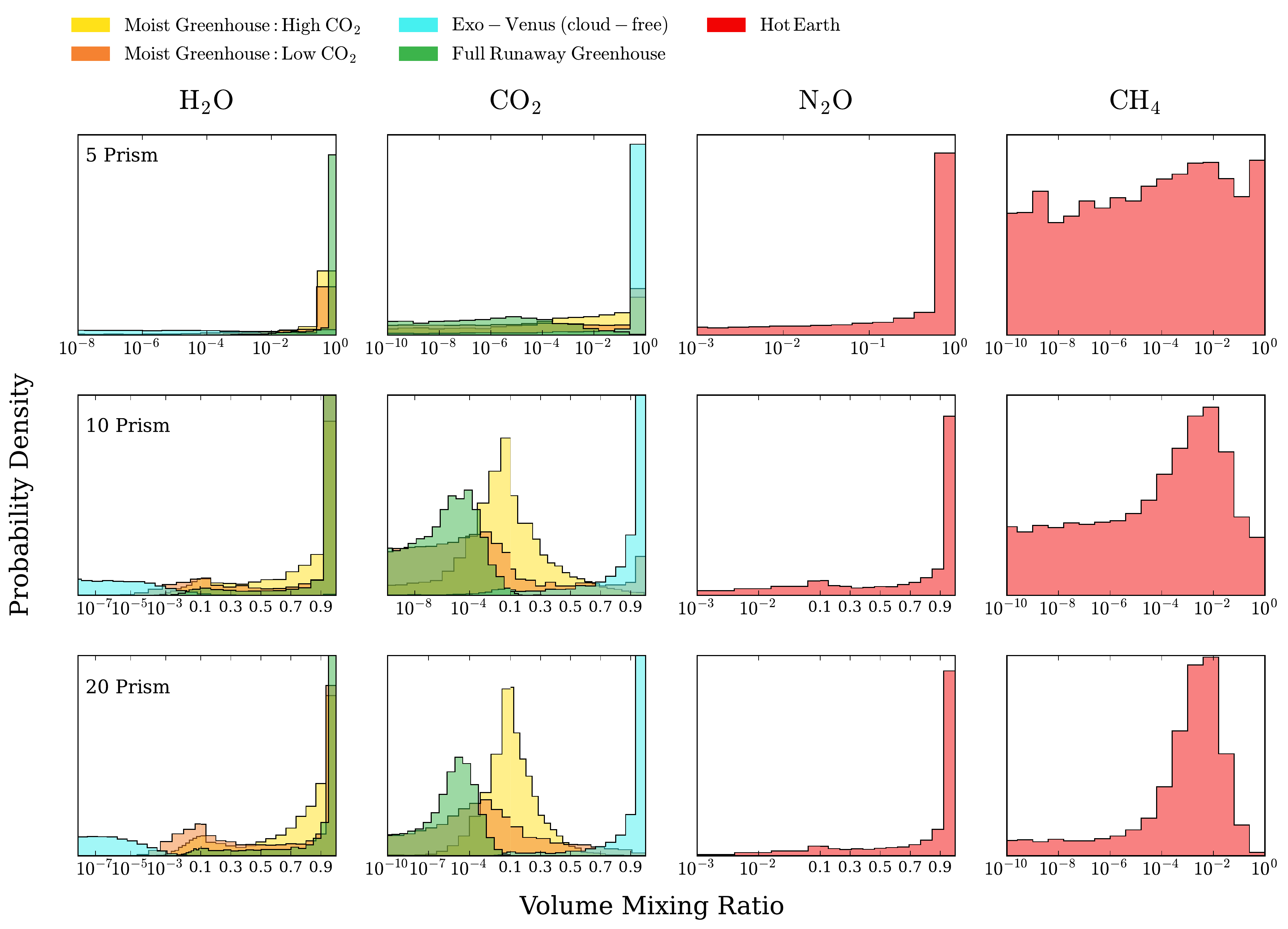}
     \vspace{-0.2cm}
    \caption{Predicted molecular abundance constraints for five potential atmospheres of LP~890-9~c. The posterior probability distributions of the molecular abundances are shown for 5, 10, and 20 \textit{JWST} NIRSpec PRISM transits (top, middle, and bottom, respectively). For ease of viewing, we omit the unconstrained $\ce{H2O}$ and $\ce{CO2}$ posteriors for the Hot Earth. We note that only the Hot Earth scenario contains $\ce{N2O}$ and $\ce{CH4}$ in the atmospheric model.}
    \label{fig:posteriors}
\end{figure*}

Figure~\ref{fig:detections} shows our predicted molecular detection significances for LP~890-9~c. The most readily detectable atmosphere arises from the Full Runaway Greenhouse scenario (Runaway 3), for which the extended $\ce{H2O}$ atmosphere produces strong $\ce{H2O}$ absorption features (see Figure~\ref{fig:retrieved_spectra}). In this Runaway 3 state, one could detect $\ce{H2O}$ to $>3\sigma$ confidence in as few as three NIRSpec PRISM transits. Our moist greenhouse scenarios require additional transits to detect $\ce{H2O}$ to $>3\sigma$: 4 transits for Runaway 2 with $20\%$ $\ce{CO2}$ abundance and 11 transits for Runaway 1 with the Earth-analog $\ce{CO2}$ abundance ($\sim 10^{-4}$). This arises from a combination of their lower atmospheric $\ce{H2O}$ abundance and/or the smaller scale height from their higher $\ce{CO2}$ abundance compared to the Full Runaway Greenhouse (Runaway 3). One can also detect $\ce{CO2}$ itself to $>3\sigma$ with 20 transits for the Runaway 2 scenario. Our cloud-free baseline Exo-Venus model, however, requires just 8 transits for a $3\sigma$ $\ce{CO2}$ detection. $\ce{CO2}$ is the most readily detectable molecule for the cloudless Exo-Venus scenario (see Figure~\ref{fig:retrieved_spectra}) due to both its prominent near-infrared absorption features \citep[e.g.][]{TheJWSTTransitingExoplanetCommunityEarlyReleaseScienceTeam2022} and its uniquely high mean molecular weight --- the latter allowing the CLR retrieval parameters to identify $\ce{CO2}$ as the bulk atmospheric consistent. However, should LP~890-9~c have a cloud coverage resembling the modern Venus, we find that 65 transits would be required to achieve the same $3\sigma$ $\ce{CO2}$ detection. Finally, the most observationally challenging scenario corresponds to LP~890-9~c being in a potentially stable habitable stage, before evolving into a Venus-like planet (our `Hot Earth' scenario). For the Hot Earth scenario, 20 transits would be required to detect $\ce{N2O}$ to $>3\sigma$. The second most detectable species in this scenario is \ce{CH4}, but with at least 40 transits for a $3\sigma$ detection. All other species ($\ce{ O2, O3, H2O, and CO2}$) in the Hot Earth model require $>100$ transits to be detected.

Figure~\ref{fig:posteriors} shows our predicted abundance constraints for \ce{H2O, CO2, N2O, and CH4} for 5, 10 and 20 NIRSpec PRISM transits. Our results indicate that 5 transits with \textit{JWST} could suffice to identify whether the main atmospheric component is $\ce{H2O}$ or $\ce{CO2}$ \citep[consistent with][]{Batalha2018}, as illustrated by the cloud-free Exo-Venus and Full Runaway Greenhouse (Runaway 3) posteriors in Figure~\ref{fig:posteriors}. Across all three runaway greenhouse models, only lower limits can be placed on the $\ce{H2O}$ abundance. With 10 transits, we predict 2$\sigma$ $\ce{H2O}$ abundance constraints of: $>$ 0.003$\%$ (Runaway 1), $>$ 0.5$\%$(Runaway 2), and $>3$$\%$(Runaway 3). We find that the best prospect for distinguishing our atmospheric scenarios arises from the retrieved $\ce{CO2}$ abundances. With 20 transits, our cloud-free Exo-Venus scenario has a strong lower limit on the $\ce{CO2}$ abundance ($>$ 59$\%$ to 2$\sigma$) while the three Runaway Greenhouse models have substantially lower retrieved $\ce{CO2}$ abundances. Runaway 2 is especially promising, with a bounded constraint on the $\ce{CO2}$ abundance (log~$\ce{CO2}$ = $-1.53^{+0.81}_{-1.65}$). Finally, our Hot Earth scenario is the most challenging to characterize due to its small spectral features (see Figure~\ref{fig:models_and_data}). With 20 transits of the Hot Earth, we find a lower limit on the $\ce{N2O}$ abundance (corresponding to the 3$\sigma$ $\ce{N2O}$ detection in Figure~\ref{fig:detections}) and a hint of $\ce{CH4}$. We provide a full table containing our predicted abundance constraints in the supplementary material.

\section{Summary and Discussion}
\label{sec:discussion}

The recently discovered super-Earth LP~890-9~c presents a rare opportunity to shed light on how the climates of rocky planets evolve at the inner edge of the habitable zone. We explored \textit{JWST}'s ability to characterize LP~890-9~c's atmosphere given a series of possible atmosphere scenarios. We find that the feasibility of characterizing the atmosphere strongly depends on the atmospheric scenario, with the most promising scenario being a Full Runaway Greenhouse scenario for which $\ce{H2O}$ could be detected to 3$\sigma$ with a small 3-transit \textit{JWST} NIRSpec PRISM program. A more ambitious 8-transit program could detect $\ce{CO2}$ for a Venus-like atmosphere without high-altitude clouds at the terminator, allowing $\ce{H2O}$-rich and $\ce{CO2}$-rich scenarios to be distinguished. Finally, a 20-transit program could even reveal N$_2$O signatures for a potentially habitable Hot Earth scenario. Observations with \textit{JWST} can therefore provide critical insights into the planetary environment. Our results suggest that, if LP~890-9~c is in the process of becoming an Exo-Venus, \textit{JWST} may be able to differentiate between our modeled evolutionary stages through which specific molecules are detected (see Figures~\ref{fig:retrieved_spectra} and \ref{fig:detections}). 

Finally, we note that some of the key assumptions in this work will benefit from further exploration in future studies. In particular, we have not fully investigated the impact of clouds or unocculted starspots on our retrieval results. Recent 3D general climate models (GCM) have demonstrated that clouds should concentrate on the night side for tidally locked rocky planets with the insolation of early Venus or early Earth \citep{turbet2021}. Therefore, clouds may not influence transmission spectra for planets like LP~890-9~c if they do not extend to the terminator region. However, further assessing the impact of clouds will require future 3D GCM models for this planet with varying cloud assumptions. Our non-inclusion of contamination from unocculted starspots (i.e. the transit light source effect; \citealt{Rackham2018}) is motivated by the low-activity of LP~890-9 \citep{Delrez2022}, which could simplify analyses of LP~890-9~c's transmission spectrum compared to the TRAPPIST-1 planets \citep[e.g.][]{Zhang2018,Wakeford2019}. However, future observations and further studies will be required to assess the relative importance of the transit light source effect for the LP~890-9 system.

Ultimately, observations of LP~890-9~c can serve to directly test both GCM predictions (by determining whether the transmission spectrum is feature-rich or featureless) and to search for signatures of stellar contamination (via short wavelength slopes). In either case, \text{JWST} observations of LP~890-9~c have the potential to provide critical new insights into our understanding of warm rocky planets that reside near the inner HZ of ultracool M-dwarfs.

\section*{Acknowledgements}
We thank the anonymous referee for a helpful report. JGB is supported by the Rose Hills Foundation. LK and JGB acknowledge support from the Brinson Foundation "Search for Life" and the Carl Sagan Institute. RJM acknowledges that support for this work was provided by NASA through the NASA Hubble Fellowship grant HST-HF2-51513.001 awarded by the Space Telescope Science Institute, which is operated by the Association of Universities for Research in Astronomy, Inc., for NASA, under contract NAS5-26555.

%%%%%%%%%%%%%%%%%%%%%%%%%%%%%%%%%%%%%%%%%%%%%%%%%%
\section*{Data Availability}

Additional figures are available in the supplementary online material \url{https://doi.org/10.5281/zenodo.7922765}.

\bibliographystyle{mnras}
\bibliography{SPECULOOS-2c.bib}

% Don't change these lines
\bsp	% typesetting comment
\label{lastpage}
\end{document}